\documentclass[twocolumn,prl,showpacs]{revtex4}

\usepackage{graphicx}
\usepackage{dcolumn}
\usepackage{float}
\usepackage{amsmath}

\newcommand{\comment}[1]{}

\begin{document}


\title{Bosonic spectral density of epitaxial thin-film La$_{1.83}$Sr$_{0.17}$CuO$_{4}$ superconductors  from infrared conductivity measurements}

\author{J. Hwang$^1$}
\email{hwangjs@mcmaster.ca}
\author{E. Schachinger$^{2}$}
\author{J.P. Carbotte$^{1,3}$}
\author{F. Gao$^4$}
\author{D.B. Tanner$^4$}
\author{T. Timusk$^{1,3}$}

\affiliation{$^1$Department of Physics and Astronomy, McMaster
University, Hamilton, Ontario L8S 4M1 Canada}
\affiliation{$^{2}$Institute of Theoretical and Computational
Physics, Graz University of Technology, A-8010 Graz, Austria}
\affiliation{$^{3}$The Canadian Institute for Advanced Research,
Toronto, Ontario M5G 1Z8, Canada} \affiliation{$^4$Department of
Physics University of Florida, Gainesville, Florida 32611, USA}

\date{\today}

\begin{abstract}
We use optical spectroscopy to investigate the excitations
responsible for the structure in the optical self-energy of thin
epitaxial films of La$_{1.83}$Sr$_{0.17}$CuO$_{4}$. Using
Eliashberg formalism to invert the optical spectra we extract the
electron-boson spectral function and find that at low temperature
it has a two component structure closely matching the spin
excitation spectrum recently measured by magnetic neutron
scattering. We contrast the temperature evolution of the
spectral density and the two-peak behavior in
La$_{2-x}$Sr$_{x}$CuO$_{4}$ with another high temperature
superconductor Bi$_{2}$Sr$_{2}$CaCu$_2$O$_{8+\delta}$. The bosonic spectral functions of the two materials account for the low $T_c$
of LSCO as compared to Bi-2212.
\end{abstract}


\maketitle

Despite twenty years of extensive research on the high critical
temperature superconducting cuprates, there is as yet no consensus
on the pairing mechanism or why $T_c$ is so high.  It is not known, for example, why most of the cuprate materials have a high $T_c$ of over 90 K while
another group, for example La$_{2-x}$Sr$_{x}$CuO$_{4}$ (LSCO) have
a much lower $T_c$ of about 30 K. Many theoretical ideas have been
put forward  such as the spin-charge separation\cite{anderson97},
preformed pairs\cite{emery95}, exchange of spin
fluctuations\cite{chubukov03,carbotte99,abanov99,zasadzinski06,campuzano99,norman98,johnson01},
and phonons\cite{lanzara01,zhou05,lee06} to list a few of the more
prominent proposals. It took more than 45 years after the
discovery of superconductivity in Hg before BCS theory was
formulated.  Yet less than ten years later, accurate and detailed
measurements of the electron-phonon spectral density
$\alpha^2F(\omega)$ responsible for the pairing were available
from tunneling experiments for many of the conventional materials.
This critical function followed from a numerical inversion,
centered on the Eliashberg equations, of the current voltage
characteristics of tunnel junctions\cite{mcmillan66}. Subsequently
optics was also used successfully to get comparable
data\cite{farnworth74,marsiglio98} and $\alpha^2F(\omega)$ was
calculated from first principles and used to calculate material
specific superconducting properties which are often quite distinct
from the universal laws of the BCS theory\cite{carbotte90}.

Many attempts have been made to obtain equivalent information in
the cuprates. An important issue in such work is whether or not
the idea of boson exchange mechanism and Eliashberg theory can in
fact be used in highly correlated systems.
Tunneling\cite{zasadzinski06,lee06}, angle-resolved
photoemission\cite{zhou05,kordyuk06,valla06}, and
optics\cite{collins89,puchkov96,tu02,dordevic05,carbotte06,schachinger06,hwang06}
have all been used to determine an electron-boson spectral density
in the oxide superconductors but with mixed results. Some argue
that the resulting spectrum is characteristic of spin fluctuations
while others attribute it to phonons. Accurate inelastic neutron
scattering measurements of the bosonic spectra of both phonons and
spin fluctuations followed by high resolution optical or tunneling
spectroscopy on the same material would go a long way towards
settling this issue\cite{chiao07}. A related technique, angle resolved photoemission (ARPES), is momentum resolved and has shown that the boson structure can depend on momentum \cite{sato07}. Thus any definitive comparison with optics would require an average of the ARPES data taken at all momenta.

Very recently inelastic neutron scattering data on the local {\it
i.e.} momentum averaged spin susceptibility in LSCO have become
available\cite{vignolle07} and show two distinct energy scales.
One peak is centered around 18 meV and the another near 40-70 meV
with small features extending to 150 meV. In this letter we invert
data\cite{gao93} on the optical conductivity of
La$_{1.83}$Sr$_{0.17}$CuO$_{4}$ within an Eliashberg formalism
with a Kubo formula for the conductivity to obtain the
electron-boson spectral density\cite{schachinger06}. We find a
remarkable similarity with the neutron data which constitutes
strong evidence for the spin fluctuation mechanism. We also
compare our new results with previous results for
Bi$_{2}$Sr$_{2}$CaCu$_2$O$_{8+\delta}$ (Bi-2212).

For optical spectroscopy La$_{2-x}$Sr$_{x}$CuO$_4$ material
presents special problems because of the presence of c-axis
longitudinal phonons in ab-plane optical spectra possibly mixed in
by an unknown lattice defect. These effects are seen in most
single crystal spectra\cite{reedyk92}. Fortunately, in epitaxial
films, these defects seem to be completely absent. We have
therefore used the low noise data from Gao {\it et al.} for our
reflectance data\cite{gao93}. The films are 820 nm thick, grown on
SrTiO$_3$ substrate by magnetron sputtering. The films show a
superconducting transition at 31 K with a transition width of 1.5
K. The details of the measurements are given in Ref.~\cite{gao93}.
\begin{figure}[t]
  \vspace*{0.0 cm}
 \includegraphics[width=9cm]{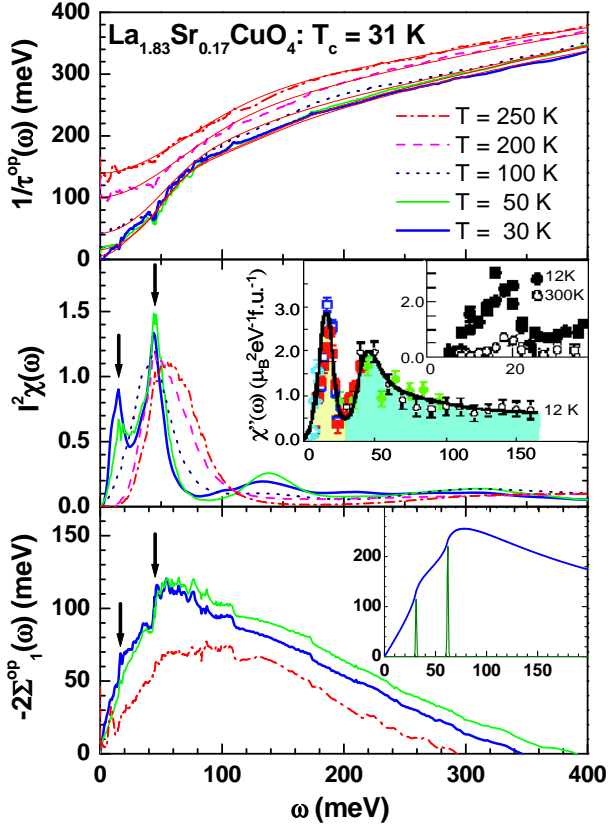}
  \vspace*{-1.0 cm}%
\caption{(color online)(Top panel) The optical scattering rate
$1/\tau^{op}(\omega)$ in meV as a function of $\omega$ (in meV)
for five temperatures. The heavy lines are the experimental data
(TO phonons have been subtracted to isolate the electronic
contribution). The light solid lines are the fits to the data.
(Middle panel) The electron-boson spectral density
$I^2\chi(\omega)$ obtained from our Eliashberg inversion from the
data in the top panel. The inset shows the data of Vignolle {\it
et al.} for a closely related sample of
La$_{2-x}$Sr$_{x}$CuO$_{4}$ with $x=0.16$ and $T_c= 38.5 $ K. The
data with solid points is at 12 K and a comparison with 300 K
data below 40 meV is also given (open points). (Bottom panel) The
real part of the optical self-energy $\Sigma^{op}_1(T,\omega)$ for
LSCO. The arrows show the positions of the sharp peaks found in
the spectral density at low temperature, $\omega = 15$  and
$\omega = 44$ meV. Note the sharp rise in the self-energy at these
frequencies. In the inset we show a simulated self-energy using a
mode with two Einstein modes.}
 \label{Fig1}
\end{figure}
The raw $\sigma(\omega)$ data for La$_{1.83}$Sr$_{0.17}$CuO$_{4}$
\cite{gao93} shows direct absorption by transverse optical phonons
as well as the electronic background of interest here. After an
appropriate subtraction of the phonons, the real and imaginary
parts of the temperature ($T$) and frequency ($\omega$) dependent
electronic conductivity $\sigma(T,\omega)$ are obtained. For
correlated electrons, a generalized Drude form applies which
defines an optical self-energy $\Sigma^{op}(T,\omega)$ through the
equation
\begin{equation}
\sigma(T,\omega)={{i\Omega_p^2} \over {4\pi}}{1 \over
{\omega-2\Sigma^{op}(T,\omega)}}
\end{equation}
where $\Omega_p$ is the plasma frequency. The optical scattering
rate $1/\tau^{op}(T,\omega)$ and the optical mass are related to
$\Sigma^{op}(T,\omega)$ by $1/\tau^{op}(T,\omega) =
-2\Sigma^{op}_2(T,\omega)$ and $\omega [m^{*op}(T,\omega)/m -1]=
-2 \Sigma^{op}_1(T,\omega)$ where 1 and 2 denote real and
imaginary parts.

Our method for determining the bosonic spectral density begins
with the application of the maximum entropy
inversion\cite{schachinger06} of a simplified convolution integral
which relates the measured scattering rate $1/\tau^{op}(T,\omega)$
to the electron-boson spectral density known to be remarkably
accurate in the normal state. This gives a first numerical model
for  $I^2\chi(\omega)$ which we then further refine\cite{hwang06}
through a least squares fit to the optical scattering rates using
the full solutions of the Eliashberg equations and associated Kubo
formula.  This gives our final model. Details about the maximum
entropy inversion and the d-wave Eliashberg equations can be found
in reference\cite{carbotte06} and \cite{schachinger06}.

In the top frame of Fig. 1 we present results for
$1/\tau^{op}(T,\omega)$ vs. $\omega$ for five temperatures. All
data are in the normal state except for the last one (30 K) which
is just below 31 K, the superconducting $T_c$  of this film. The
heavy curves correspond to the data which is used as the input in
the inversion process. Our final results for the electron-boson
spectral density $I^2\chi(\omega)$ are shown in the middle panel.
Note the strong temperature dependence of the low energy structure.
When this $I^2\chi(\omega)$ function is used in the
Eliashberg equations, the optical scattering rates that result are
shown in the top frame as the light lines. In all cases the fit to
the data is good. For the 30 K data in the superconducting state
d-wave symmetry was assumed for the gap channel.

The middle frame of Fig. 1 shows the extracted bosonic spectral
density. Note that at 30 and 50 K both spectra show a two-peak
structure with peaks at 15 and 44 meV. As the temperature is
increased the spectral density evolves towards a single peak which
broadens and moves to higher energies. We also note tails
extending beyond 100 meV but these have a low amplitude as
compared to the peaks. All these features are in agreement with
the data on the local spin susceptibility (shown in the inset)
measured by polarized neutron scattering in a closely related
sample by Vignolle {\it et al.}\cite{vignolle07}. The data at 12 K
shows the same two peak structure as we have found at 30 K. While
Vignolle {\it et al.} have limited data at 300 K they are also in
accord with the temperature dependence of our spectra: the low
energy peak in $I^2\chi(\omega)$, which exists only at low
temperature is absent above 100 K as it is in the neutron spectra
at 300 K. The detailed agreement between $I^2\chi(\omega)$ and the
neutron data on the local magnetic susceptibility $\chi''(\omega)$
is strong evidence that the charge carriers in
La$_{1.83}$Sr$_{0.17}$CuO$_{4}$ are coupled through a spin
fluctuation mechanism. In related ARPES work on highly underdoped LSCO, Zhou {\it et al.}\cite{zhou05}  interpret their results in terms of phonons.  We note that the magnetic resonance peak is known to be very weak in highly underdoped materials\cite{yamani07} and that (as discussed above) nodal direction ARPES cannot be directly compared with momentum averaged optics.

A phenomenological model for the oxides where spin fluctuations
replace the usual phonon exchange is the nearly antiferromagnetic
fermi liquid model (NAFFL)\cite{chubukov03}. This model is well
developed and is anchored in the generalized Eliashberg equations. While our approach to the inversion of the
optical data might not apply in some highly correlated metals, it
is fully justified in the NAFFL model. In particular for a
discussion of the applicability of Migdal's theorem (a
precondition in our approach) we refer the reader to the review by
Chubukov {\it et al.}\cite{chubukov03}.

Two peak structure in the fluctuation spectrum of
La$_{1.83}$Sr$_{0.17}$CuO$_{4}$ at low temperature and the
disappearance of the lower peak at high temperatures can be seen
directly in the optical self-energy itself which follows from the
raw data without any appeal to microscopic models. This can be
seen clearly in the bottom panel of Fig. 1 where we show our data
for $-2\Sigma^{op}_1(T,\omega)$ for three temperatures, T = 30 K
in the superconducting state (just below $T_c=31$ K), at 50 K,
above $T_c$, and 250 K. The heavy black arrows indicate the
position of the two peaks in $I^2\chi(\omega)$. They coincide with
sharp rises in $-2\Sigma^{op}_1(T,\omega)$. Such sharp rises are
expected in the normal state for coupling of electrons to two
Einstein oscillators as shown for a model calculation in the inset
where a model spectrum $I^2\chi(\omega)$ has two peaks at 31 and
62 meV of width 1.2 meV with the second peak having twice the
spectral weight of the first. The two peaks in $I^2\chi(\omega)$
can also be seen directly in the second derivative $2\pi
W(\omega)=d^2[\omega/\tau^{op}(\omega)]/d\omega^2$ (not shown
here) which is a model independent method and is known to be
closely related to the spectral density in the peak
region\cite{marsiglio98}. The red dashed-dot curve at $T=250$ K
shows a single rise consistent with a single peak in
$I^2\chi(\omega)$ at this temperature.
\begin{figure}[t]
  \vspace*{-1.0 cm}
 \includegraphics[width=9cm]{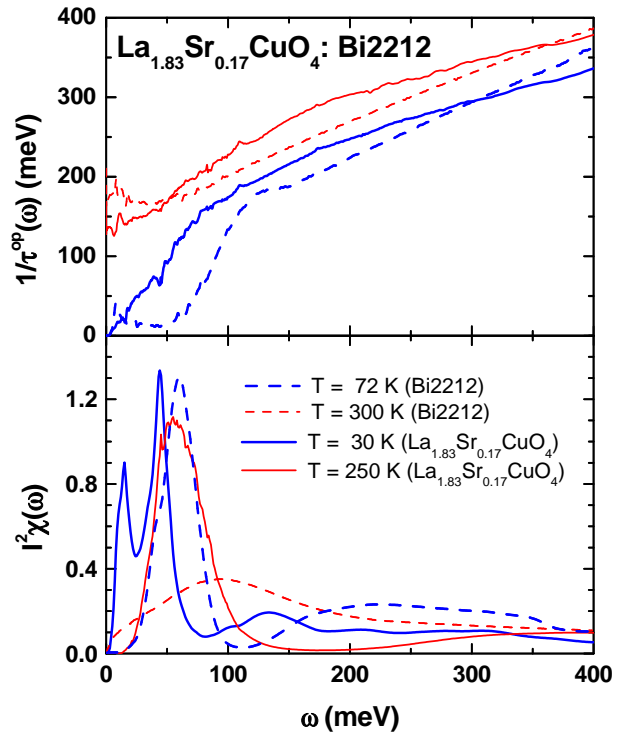}
  \vspace*{-1.5 cm}%
\caption{(color online) (Top panel) Scattering rates for Bi-2212
(dashed lines) and LSCO (solid lines). The two materials show
striking differences at low energies. Scattering sets in at very
low energies in the LSCO material whereas Bi-2212 has a negligible
scattering below 50 meV. (Lower panel) The bosonic spectral
function $I^2\chi(\omega)$ of the same materials shown in the top
panel. At low temperature LSCO has two peaks while Bi-2212 shows a
single peak at much higher energy. As the temperature is raised
the Bi-2212 peak is weakened dramatically and moves to high
frequency~\cite{hwang06} whereas in LSCO the lower peak vanishes
while the upper peak remains up to 250 K. This difference at high
temperature can already be seen in the top panel: the LSCO sample
shows an onset of scattering in the 50 meV region while the Bi-2212
curve is quite featureless at 300 K.}
 \label{Fig2}
\end{figure}

It is interesting to compare LSCO with similar results for
optimally doped Bi-2212\cite{hwang06} ($T_c=96$ K). These Bi-2212
data are shown in Fig. 2. The optical scattering rates on which
our inversions are based are shown in the top frame while the
lower frame shows our results for the spectral densities
$I^2\chi(\omega)$. Only two temperatures are shown 72 K and 300 K
for Bi-2212 (dashed curves) and 30 and 250 K for LSCO (solid
curves). The scattering rates are very different in the two
materials. The dashed curve for Bi-2212 shows a very steep rise
with the mid point at 92 meV which corresponds to the energy as
the peak in the real part of the optical self-energy. By contrast
the rise in the thick solid blue curve for LSCO is more gradual
and proceeds on a broader energy scale. The rise starts close to
zero energy because of the low energy peak in $I^2\chi(\omega)$
(thick solid blue curve in the bottom panel) and is much broader
in energy because of the second peak around 50 meV. At higher
temperatures the thin solid and the thin dashed red curves still
show distinct variations with $\omega$. For Bi-2122 the underlying
spectrum $I^2\chi(\omega)$ has evolved to a single broad peak at
about 100 meV plus a background. For LSCO the spectrum shows a
sharper peak at about 55 meV and the background in comparison is
small. This translates into a room temperature scattering rate in
Bi-2212 which is featureless and flat while in LSCO the bosonic
peak is very much present giving rise to a marked shoulder in the
scattering rate in the 50 meV region.

What are the implications of these spectra for superconductivity?
$T_c$ in Bi-2212 is three times larger than
in LSCO. For an s-wave electron-phonon superconductor the same
spectral density $I^2\chi(\omega)$ enters both renormalization and
gap channels of the Eliashberg equations. In a d-wave superconductor,
however, the spectral density that enters the gap channel is a
d-wave projection of the electron-boson exchange process rather
than its s-wave projection which we have determined from optics
and shown in the middle frame of Fig. 1 and the lower frame of Fig. 2.
For simplicity we can assume that these two quantities differ
mainly by a numerical factor\cite{carbotte03}. This factor can be
fit to the known value of the critical temperature $T_c$ which we
determine from the complete numerical solutions of the d-wave
Eliashberg equations. We find that $T_c$ is well represented by
the simple modification of the McMillan
equation\cite{mcmillan68,williams89}
\begin{equation}
k_B T_c \cong \hbar\:
\omega_{ln}\exp\Big{[}-\frac{1+\lambda^s}{\lambda^d}\Big{]}
\end{equation}
where $1+\lambda^s$ comes from the normal state renormalization of the dispersion curves and $\lambda^d$ is the interaction in the gap channel.  Here $\omega_{ln}$ is the
average boson energy defined by Allen and Dynes\cite{allen75} and is the same for
s- or  d-channel cases.

Using our $I^2\chi(\omega)$ functions we find that the $\lambda^d$ values are nearly the same for Bi-2212
(1.85) and LSCO (1.90). However the value of $\omega_{ln}$ differ
by a factor of two, $\sim$50 meV for Bi-2212 and $\sim$25 meV for
LSCO. Therefore the softening of the spin fluctuation spectrum in
LSCO as compared to Bi-2212 accounts for a factor of 2 difference
in $T_c$. The remaining difference is traced to the value of
$\lambda^s$ which is larger in LSCO (3.40) as compared to Bi-2212
(2.50). The renormalization factor
$1+\lambda^s$ in the modified McMillan equation (Eq. (2)) is pair
breaking and this accounts for the rest of the difference in $T_c$
values. While we do not predict $T_c$ we can explain in a robust way
from our analysis the factor of 3 difference in $T_c$ between LSCO
and Bi-2212 samples.

In summary we have measured the fluctuation spectrum in LSCO,
$I^2\chi(\omega)$ at various temperatures. At low temperature it
shows two characteristic energy scales in remarkable agreement
with the local (q averaged) spin susceptibility recently found in
polarized inelastic neutron scattering experiments. As the
temperature is increased, the low energy peak disappears in accord
with the neutron results. While the maximum entropy technique and
least squares fit to the measured optical scattering rates within
an Eliashberg framework is employed here, the two-peak structure
in $I^2\chi(\omega)$ can be seen directly in the raw data for the
real part of the optical self-energy. In contrast optimally doped Bi-2212 reveals a
very different behavior showing a single sharp peak centered at 60
meV with a valley above it and a broad low intensity background
extending to energies up to 400 meV. Finally, the bosonic spectra
derived from our analysis, fully
account for the low superconducting transition temperature of LSCO
as compared to Bi-2212.

\centerline{\bf Acknowledgments}

This work has been supported by the Natural Science and
Engineering Research Council of Canada and the Canadian Institute
for Advanced Research. FG and DBT acknowledge support from the NSF
and DOE through grants DMR-0305043 and DE-AI02-03ER46070." Also 
we want to acknowldege J. Talvacchio and M.G. Forrester for
preparing the film.


\begin{thebibliography}{99}
\bibitem{anderson97} P. W. Anderson, The theory of superconductivity in the high-T$_c$
cuprates. Princeton University Press, Princeton, New Jersey (1997).

\bibitem{emery95} V. J. Emery and S. A. Kivelson, Nature (London) {\bf 374}, 434  (1995).


\bibitem{chubukov03} A. V. Chubukov {\it et al.},
The Physics of Conventional and Unconventional Superconductors
edited by K.H. Bennemann and J.B. Ketterson, Springer-Verlag
(2003).

\bibitem{carbotte99} J. P. Carbotte, E. Schachinger and D. N. Basov, Nature (London) {\bf 401}, 354 (1999).
\bibitem{abanov99} Ar. Abanov and A. V. Chubukov, Phys. Rev. Lett. {\bf 83}, 1652 (1999).

\bibitem{zasadzinski06} J. F. Zasadzinski {\it et al.}, Phys. Rev. Lett. {\bf 96}, 017004
(2006).

\bibitem{campuzano99} J. C. Campuzano {\it et al.}, Phys. Rev. Lett. {\bf
83}, 3709 (1999).


\bibitem{norman98} M. R. Norman and H. Ding, Phys. Rev. B {\bf 57}, R11089 (1998).

\bibitem{johnson01}  P. D. Johnson {\it et al.}, Phys. Rev. Lett. {\bf 87}, 177007 (2001).
\bibitem{lanzara01} A. Lanzara {\it et al}, Nature
(London) {\bf 412}, 510 (2001).

\bibitem{zhou05} X. J. Zhou {\it et al.}, Phys. Rev. Lett. {\bf 95}, 117001
(2005).

\bibitem{lee06} J. Lee {\it et al.}, Nature (London) {\bf 442}, 546 (2006).

\bibitem{mcmillan66}  W. L. McMillan and J. M. Rowell, Phys. Rev. Lett. {\bf 14}, 108 (1966).


\bibitem{farnworth74} B. Farnworth and T. Timusk, Phys. Rev. B {\bf 14},
2799 (1974).
\bibitem{marsiglio98} F. Marsiglio, T. Startseva and J. P. Carbotte, Phys. Lett. A {\bf 245}, 172 (1998).

\bibitem{carbotte90} J. P. Carbotte, Rev. Mod. Phys. {\bf 62}, 1027 (1990).

\bibitem{kordyuk06}  A. A. Kordyuk {\it et al.}, Phys. Rev. Lett. {\bf 97}, 017002 (2006).

\bibitem{valla06} T. Valla {\it et al.}, cond-mat/0610249 (2006).


\bibitem{collins89} R. T. Collins {\it et al.}, Phys. Rev B {\bf 39,} 6571 (1989).

\bibitem{puchkov96}  A. V. Puchkov, D.N. Basov and T. Timusk,
J. Physics, Condensed Matter {\bf 8}, 10049 (1996).

\bibitem{tu02}  J. J. Tu {\it et al.}, Phys. Rev. B {\bf 66},
144514 (2002).

\bibitem{dordevic05} S. V. Dordevic {\it et al.}, Phys. Rev. B {\bf 71}, 104529 (2005).

\bibitem{carbotte06} J. P. Carbotte and E. Schachinger, Ann. Phys. {\bf 15}, 585 (2006).

\bibitem{schachinger06} E. Schachinger, D. Neuber and J. P. Carbotte, Phys. Rev. B {\bf 73}, 184507 (2006).
\bibitem{hwang06} J. Hwang, T. Timusk, E. Schachinger and J. P. Carbotte, Phys. Rev. B {\bf 75}, 144508 (2007).

\bibitem{chiao07} M. Chiao, Nature Physics {\bf 3}, 148 (2007).

\bibitem{sato07} T. Sato {\it et al.}, J. Phys. Soc. Japan, {\bf 76}, 103707 (2007).


\bibitem{vignolle07} B. Vignolle {\it et al.}, Nature Physics {\bf 3}, 163 (2007).

\bibitem{gao93} F. Gao {\it et al.}, Phys. Rev. B {\bf 47}, 1036 (1993).

\bibitem{reedyk92} M. Reedyk and T. Timusk, Phys. Rev. Lett. {\bf 69,} 2705 (1992).

\bibitem{yamani07} Z. Yamani {\it et al.} cond-mat/0706.0327.

\bibitem{carbotte03} J. P. Carbotte and E. Schachinger, Model and
Methods of high-$T_c$ Superconductivity Edited by J. K. Srivastava
and S. M. Rao, Nova Science Publishers Inc., Hauppauge N.Y. p 73 (2003).

\bibitem{mcmillan68} W. L. McMillan, Phys. Rev. {\bf 167}, 331
(1968).

\bibitem{williams89} P. J. Williams and J. P. Carbotte, Phys. Rev.
B {\bf 39}, 2180 (1989).

\bibitem{allen75} P. B. Allen and R. C. Dynes, Phys. Rev. B {\bf
12}, 905 (1975).

\end{thebibliography}

 \end{document}